# Advanced optical designs of curved detectors-based two-mirrors unobscured telescopes

Eduard R. Muslimov[a,b*], Emmanuel Hugot[a], Simona Lombardo[a], Melanie Roulet[a],
Thibault Behaghel[a], Marc Ferrari[a], Wilfried Jahn[c];

[a]Aix Marseille Univ, CNRS, CNES, LAM, Marseille, France, 38, rue Joliot-Curie, Marseille,13388, France; [b]Kazan National Research Technical University named after A.N. Tupolev –KAI, 10 K. Marx, Kazan 420111, Russia; [c] California Institute of Technology, 1200 E. California Blvd, Pasadena, California 91125, USA.

## ABSTRACT

In the present paper we consider a family of unobscured telescope designs with curved detectors. They are based on classical two-mirror schemes – Ritchey-Chretien, Gregorian and Couder telescopes. It is shown that all the designs provide nearly diffraction limited image quality in the visible domain for 0.4º×0.4º field of view with the f-number of 7. We also provide a brief ghost analysis and point on special features of the systems with curved detectors. Finally, the detector surface shape obtained in each case is analyzed and its' technological feasibility is demonstrated.

## 1. INTRODUCTION

Optical systems of unobscured telescopes have a number of obvious advantages like increase of the primary collecting area or exclusion of the diffraction on spiders. The simplest way to generate an unobscured design consists of the pupil shift or the field of view bias in an ordinary coaxial telescope design. Such a transformation for classical two-mirror designs was considered many times[1,2]. Since the scheme symmetry in such an optical design is broken, some authors[2] propose to add more degrees of freedom by tilting the components in order to provide a high optical quality at high *f*-number values.

On the other hand, it is known that one of the main factors limiting an off-axis use of the classical telescopes is the field curvature[3]. However, the recent progress in curved detectors technology[4-10] allows to compensate the field curvature by the detector shape. It means that the condition of the field curvature correction can be neglected when compensating the system's aberrations. It was shown by a number of authors[9,10] that using of a curved detector changes the optical design approach and makes it possible to reach a better image quality along with simplification of the optical design and increasing of the image illumination uniformity.

Thus the main goal of the present study is to demonstrate the prospective advantages of use of curved detectors in off-axis unobscured two-mirror telescopes. We consider three optical designs based on such well-known schemes as Ritchey-Chretien, Gregorian and Couder telescopes. In Section 2 we present the design algorithm applied for each of the schemes to produce an advances unobscured version. Section 3 presents analysis of the unobscured designs geometry and optical quality. In Section 4 the detector surface shape obtained in each case is analyzed. Section 5 contains the general conclusion on the study.

## 2. DESIGN PROCEDURE

For each of the considered two-mirror telescope schemes the following re-design procedure is used:

1. The initial optical scheme is scaled for the necessary focal length.
2. For the initial design the theoretical value of field curvature radius is calculated[11]. In this particular case we assume that the radius corresponds to image surface of minimum wavefront error:

---

[*] eduard.muslimov@lam.fr; phone +33 4 91 05 69 18; lam.fr

$$R_I \approx -D_p F_p \frac{F_p + \eta}{F_p + F - \eta} \quad (1)$$

Where $D_p$ is the primary's diameter, $F_p$ and $F$ are the primary's and overall *f*-ratio values, respectively, and $\eta$ is ratio of the back focal length to the primary's diameter. The value found according to (1) is substituted as the image surface curvature radius.

3. The entrance pupil is shifted by a distance allowing to exclude any central obscuration.

4. The entire system is numerically optimized with the standard tools implemented in Zemax software. The curvature radii (including the one of the detector surface), the conic constants and the distances are used as variables. The boundary conditions require maintenance of the focal distance and the principal geometry.

5. Tilt angle of each of the components is introduced and used as a variable parameter. The aperture off-axis shift is also set as a variable. The optimization is repeated. The additional boundary conditions on the marginal rays coordinates differences are used to avoid obscuration.

Below we present the designs derived from classical two-mirror systems by using of these procedure.

## 3. OPTICAL DESIGNS OVERVIEW

The design procedure described above was applied to three classical two-mirror telescope designs, namely Ritchey-Chretien, Gregorian and Couder schemes. For simplicity of comparison the principal optical parameters were the same in all the cases. They are presented in Table 1.

Table 1. Principal parameters of the telescopes optical designs.

| Focal length, mm | 1500 |
|---|---|
| F-number | 7 |
| Field of view, ° | 0.4x0.4 |
| Optical scheme type | Two-mirror unobscured off-axis telescope |
| Detector surface | Spherical |

### 3.1 Optical schemes

All the optical schemes at different design steps are shown on Figure 1. Line *a* corresponds to the initial scheme (see p.1), line *b* corresponds to the off-axis design with curved detectors (see p.2-4), and line *c* represents the finalized design with tilted components and the real apertures. Note, that all the images have different scales. The actual dimensions for the finalized designs are: 216×330×438 mm³ (X×Y×Z) for Ritchey-Chretien, 216×293×515 mm³ for Gregorian, 216×429×1715 mm³ for Couder. Though in the Couder scheme we have to use an additional folding flat mirror to limit the telescope length. One can note that the components tilt angles in the Ritchey-Chretien type and Couder type designs are significant, while in the Gregorian telescope the components remain on the same axis even after the re-optimization.

The designed schemes, especially the Ritchey-Chretien based one, are interested in terms of ghost analysis. The result of non-sequential raytracing for this system is shown on Figure 2. On one hand, the Ritchey-Chretien type design keeps such an advantage as possibility of simple internal baffling. If two simple screens are set next to the primary and secondary mirrors, the minimum angle of a ray passing through them is 13.8° and it falls outside of the detector sensitive area. The other beams from side sources are blocked. On the other hand, if we account for the detector surface reflectivity, a part light will be reflected back to the system and will reach the secondary (see the rays shown in yellow and violet on Figure 2). One of the reason why this undesired effect occurs is the influence of the detector shape on the reflected beam geometry. This parameter wasn't directly controlled during the optimization and it is still possible to read of this back reflection by means of the tilt angles adjustment. However, this simple analysis indicates an additional feature of the curved detectors. So an additional attention should be paid to the ghost analysis in a system with curved detectors.

A similar analysis was performed for the two other designs. In the Couder type system the light reflected back from the detector doesn't fall onto the secondary mirror due to the sign of the tilt angle. Also, in this case the folding mirror excludes possibility of a direct illumination of the detector. In addition, all the mirrors operate with relatively slow *f*-numbers, so there is enough space for internal baffles. In the Gregorian system the beam reflected back from the detector is partially blocked by the rear side of primary mirror, but a portion of it also can be re-reflected from the secondary. An internal baffling in the Gregorian type design is impossible as one can see from Figure 1.

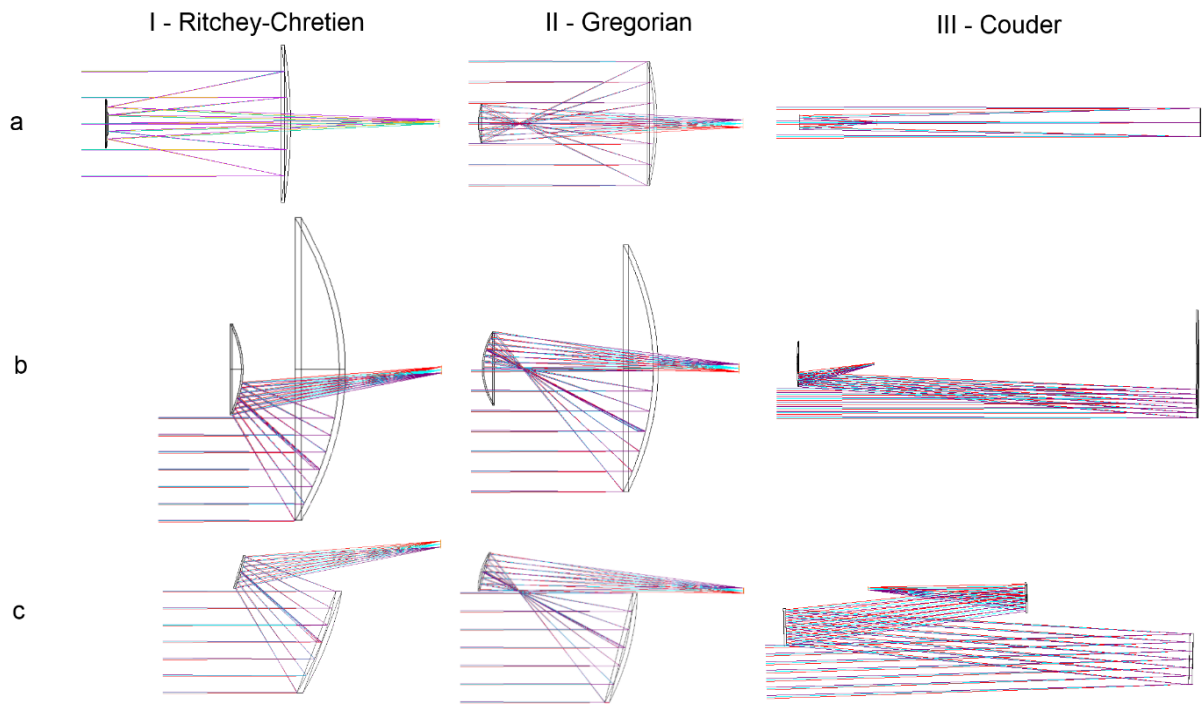

Figure 1. Generation of the unobscured two-mirror telescope designs with curved detectors: a – initial designs, b – off-axis configurations with curved detectors, c – finalized designs.

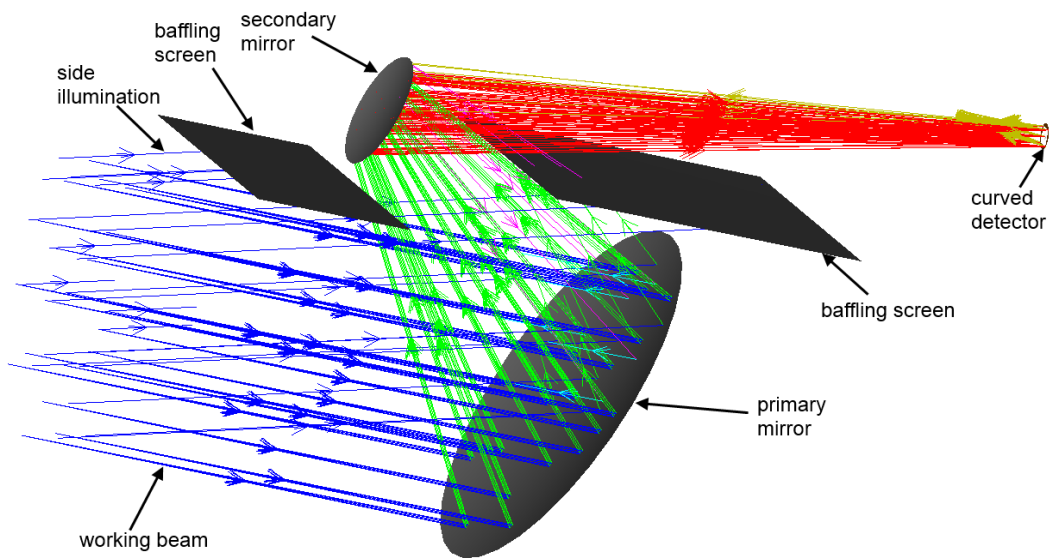

Figure 2. Non-sequential raytracing for the ghost analysis in the Ritchey-Chretien type scheme.

## 3.2 Image quality

For estimation of the image quality standard spot diagrams are used. Since the optical schemes have no axial symmetry we consider a rectangular grid covering the square field of view. The reference value is the Airy disk radius, which is the same for all the presented designs and equal to 4.7 µm at wavelength of 550 nm.

The spot diagrams for Ritchey-Chretien type design are shown on Figure 3. The root-mean square (RMS) radii vary between 0.9 and 3.5 µm. The image quality is close to the diffraction limit for the entire field, though the image quality degrades rapidly towards the field corners. One also can note absence of a symmetry with respect to the XZ plane.

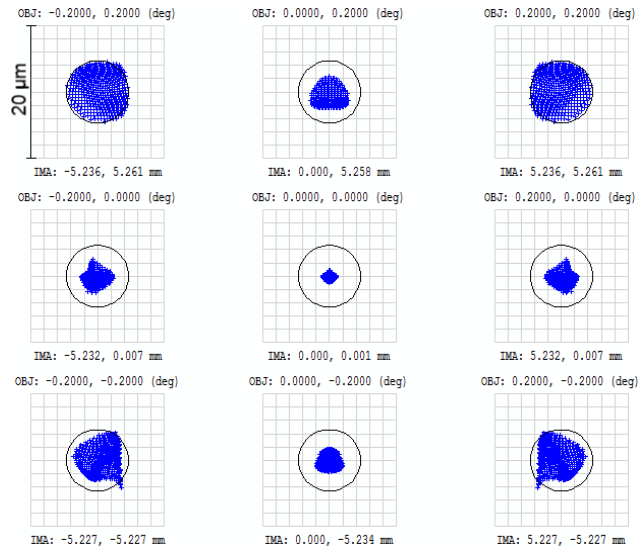

Figure 3. Spot diagrams of the Ritchey-Chretien type telescope design.

In a similar way the spot diagrams for Gregorian type design are shown on Figure 4. The RMS radii are 0.1-3.4 µm. The spots burring at the field of view corners is even more notable, though the system remains diffraction limited. Since the aperture offset is smaller and the components tilt angles are negligible, the spot size distribution is more symmetrical.

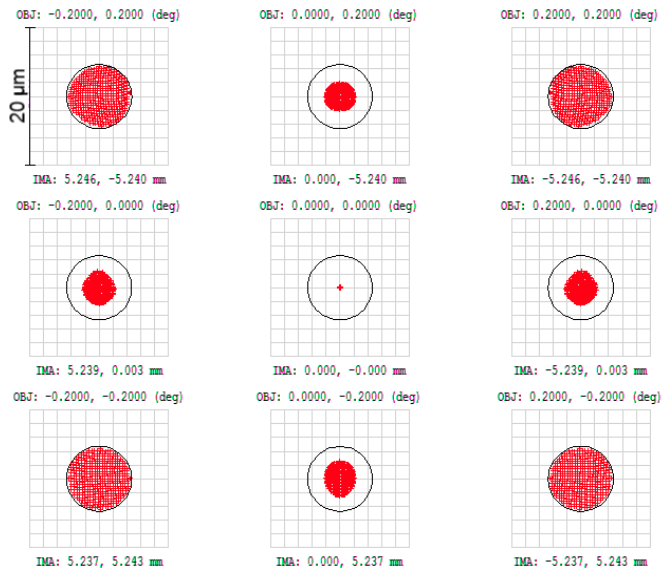

Figure 4. Spot diagrams of the Gregorian type telescope design.

Finally, the spot diagrams for the Couder based design are shown on Figure 5. The RMS radius values are 1.9-3.1 µm. The spots radii distribution is more uniform, but the spots sizes exceed the Airy disk diameter because of uncompensated aberrations. So the telescope cannot be literally defined as a diffraction limited one, though it approaches the limit.

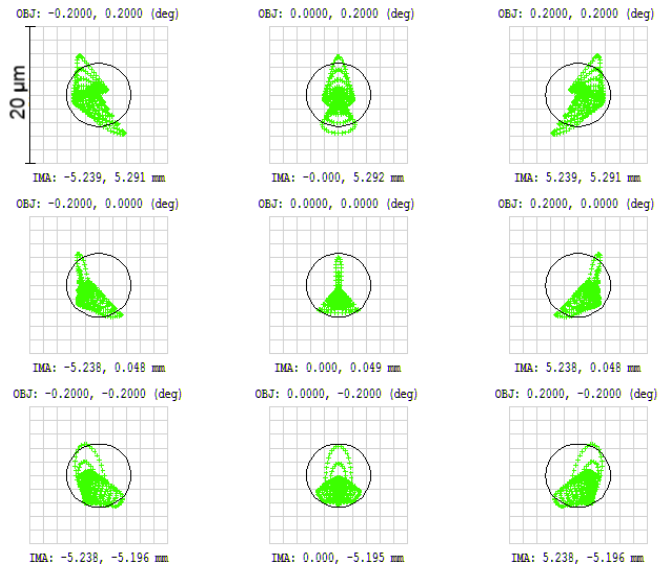

Figure 5. Spot diagrams of the Couder type telescope design.

The spot diagrams data is summarized in table 2. For comparison we provide also the values of spot sizes for a case of a flat detector. In this case the detector surface curvature was fixed at zero and the entire system was re-optimized. One can see that for the Ritchey-Chretien and Gregorian types the difference in the spot radii can be as big as an order of magnitude. It emphasizes the importance of use of the curved detector in such a scheme. For the Couder type design the image surface steepness is much less, so the difference equals only to 13% in the worst case.

Table 2. Spot diagrams data.

|  | **Ritchey-Chretien** | **Gregorian** | **Couder** |
|---|---|---|---|
| On a curved surface | | | |
| Spot RMS radius in center/corner, µm | 0.9/3.5 | 0.1/3.4 | 1.7/2.9 |
| Spot MAX radius in center/corner, µm | 1.5/5.4 | 0.2/4.8 | 5.3/7.7 |
| On a plane | | | |
| Spot RMS radius in center/corner, µm | 10.1/7.6 | 17.1/6.6 | 1.9/3.0 |
| Spot MAX radius in center/corner, µm | 14.5/14.3 | 22.7/12.3 | 6.0/7.0 |

### 4. CURVED DETECTOR DESIGN

Since the focal length and the angular field of view is the same for all the considered designs, the detector dimensions are the same in all the cases and equal to 10.5×10.5 mm$^2$. The sag map for each of the curved detectors is shown on Figure 6.

The key values defining the detector surface shape are also given in Table 3 below. Note that the radii values found after optimization are close to those computed analytically for the initial configurations.

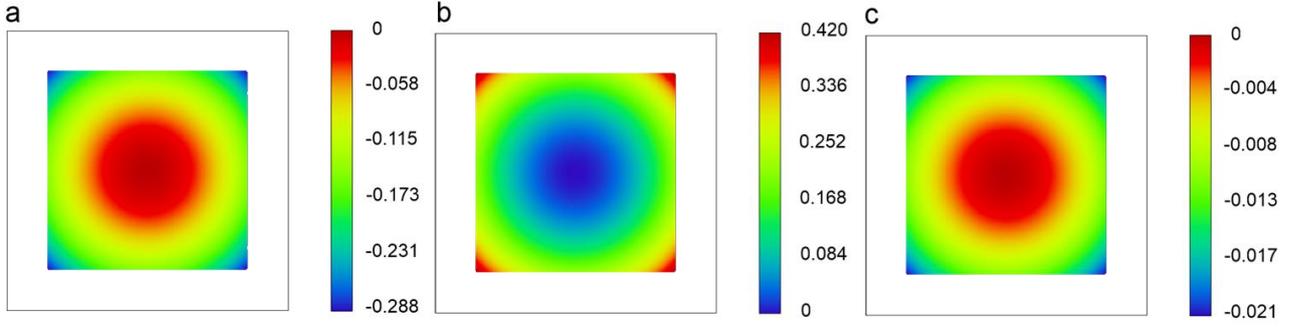

Figure 6. Sag maps of the curved detector surfaces: a – Ritchey-Chretien type design, b – Gregorian type design, c – Couder type design.

Such a spherical shape can be generated by mechanical bending of a back-thinned detector, attached on top of a thin metallic plate[12,13]. This technology represents an extension of the active deformable mirrors technology, described in a number of sources[14,15]. This technology allows to bend a thin polished plate to the desired shape using a non-uniform load or a special distribution of the plate thickness. It was proven that with the same approach a commercial CMOS or CCD detector can be bent to the required radius of curvature and thus a fully functional curved detector can be manufactured.

Here we do not provide a detailed analysis of the calculated detectors shapes feasibility. Instead we make an approximate theoretical estimation of the maximum stress generated in the bending setup.

We assume that a thin silicon chip (a typical thickness is 100μm) is glued firmly on top of a circular metallic plate, so the chip almost doesn't affect the rigidity of the entire assembly and the stress and deformation distributions don't have discontinuities at the chip-to-plate interface. The plate diameter is equal to the detector's diagonal. Then we consider the simplest case of bending of the plate with supported edge by a uniform load. It may produce nearly spherical shape[16].

The maximum sag generated in such a bending setup is defined as

$$w_{max} = \frac{(5+\nu)qa^4}{64(1+\nu)D}, \quad (2)$$

where $\nu$ is the Poisson ratio, $q$ is the uniform load, $a$ is the plate radius and $D$ is the flexural rigidity defined as

$$D = \frac{Eh^3}{12(1-\nu^2)}. \quad (3)$$

Here $h$ is the plate thickness and $E$ is the Young modulus.

The maximum stress is

$$\sigma_{max} = \frac{3(3+\nu)qa^2}{8h^2}. \quad (4)$$

Simply substituting the eq. (2) and (3) into (4) we can define the maximum stress for a given plate thickness and the required sag, i.e. to the required curvature. Let us assume that the bending facility is made of an Al alloy having $\nu$=0.33 and $E$=69GPa. The results of computation with this simple analytical model for the detector curvature and linear field values taken from the optical designs are shown on Figure 7. The maximum stress values for a 1mm-plate are also given in Table 3.

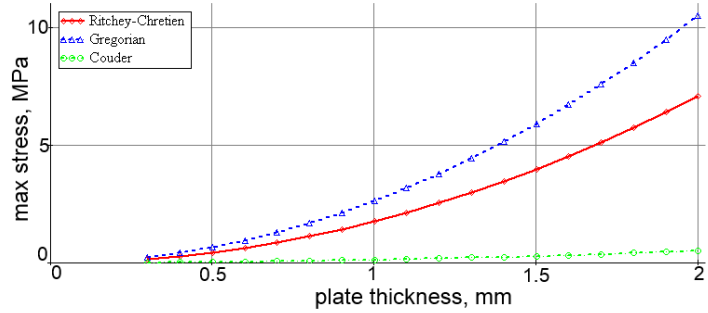

Figure 7. Estimation of the maximum stress generated during the detector bending.

The maximum generated stress is much less than the tensile strength of silicon, which equals to 350 MPa. Obviously, this is an approximate estimation demonstrating that there is no risk of breaking the detector during its bending. A more precise modelling requires use of finite element analysis (FEA). Such a detailed study was carried out before for a small curved detector[13].

Table 3. Curved detectors data

|  | **Ritchey-Chretien** | **Gregorian** | **Couder** |
|---|---|---|---|
| Initial radius of curvature, mm | -125.6 | 59.5 | -1358.9 |
| Optimized radius of curvature, mm | -95.8 | 65.6 | -1314.3 |
| Sag, mm | 0.29 | 0.43 | 0.02 |
| Maximum stress, MPa | 1.77 | 2.62 | 0.13 |

## 5. CONCLUSIONS

In the present paper we demonstrated use of curved detectors in optical schemes of two-mirror unobscured telescopes. Each of the schemes is derived from a classical design (Ritchey-Chretien, Gregorian and Couder) by re-optimization with off-axis aperture, curved image plane and tilted components. It was shown that all the schemes provide nearly diffractive image quality over 0.4º×0.4º field with the *f*-number of 7. They also keep some advantages inherent for a specific geometry, like a possibility of internal baffling for the Ritchey-Chretien off-axis geometry. Meanwhile, it was found that a special attention should be paid to the ghosts' analysis if a curved detector is used.

We also provided a brief analysis of the detectors shapes and their feasibility. It is shown that the required curvature can be produced without generating any dangerous mechanical stress. We assume that such a simplified analytical estimation can be extremely useful in design of new optical schemes with curved detectors, because it can be used to define a boundary condition directly in the optical system optimization procedure. In some cases, it may be more convenient than use of precise calibration curves confirmed by FEA and experiments.

In general, the curved detector technology and the new optical designs, which becomes possible due to its use, are of interest for a number of application fields. However, the primary application of the curved detector should be optical devices for scientific research, especially the future astronomical instruments[18-20], which require a perfect performance and operate in unusual modes.

## ACKNOWLEDGEMENTS

The authors acknowledge the support from the European Research council through the H2020 - ERCSTG-2015 - 678777 ICARUS program. This research was partially supported by the HARMONI instrument consortium. We also thank Christophe Gaschet, Bertrand Chambion and David Henry from Univ. Grenoble Alpes, CEA-LETI for their contribution to the curved detectors technology development.